\documentclass[a4paper,twosides]{article}
\usepackage[utf8]{inputenc}
\usepackage{iftex}
\ifPDFTeX
  \usepackage{CJK}
\fi
\usepackage{multicol,multirow,graphicx,fancyhdr,epstopdf}
\usepackage{amsmath,amsfonts,amssymb,amsthm,bm,upgreek,mathrsfs,mathcomp}
\IfFileExists{ccmap.sty}{\usepackage{ccmap}}{}
\usepackage[pagewise,switch,columnwise]{lineno}
\usepackage[compress,nospace]{cite}
\usepackage[dvipsnames]{xcolor}
\usepackage{caption}
\usepackage{subcaption}
\usepackage{booktabs}
\usepackage{array}
\usepackage{siunitx}
\usepackage{listings}
\usepackage[ruled,vlined,linesnumbered]{algorithm2e}
\usepackage{url}
\usepackage{CPL-2023}

\renewcommand{\arraystretch}{1.1}

\newcommand{\cplyear}{xxxx} \newcommand{\cplvol}{xx}
\newcommand{\cplno}{x} \newcommand{\cplpagenumber}{xxxxxx}
 \newcommand{\cplpage}{\cplpagenumber-\thepage}

\newcommand{\BECTensor}[9]{%
  \ensuremath{%
    \left[
    \vcenter{\hbox{%
        \begin{tabular}{@{}S[table-format=-1.3]S[table-format=-1.3]S[table-format=-1.3]@{}}
          #1 & #2 & #3 \\
          #4 & #5 & #6 \\
          #7 & #8 & #9
        \end{tabular}%
    }}
    \right]
  }%
}
\SetKwComment{Comment}{$\triangleright$\ }{}

\ifPDFTeX
  \newcommand{\CPLbeginCJK}{\begin{CJK}{GBK}{song}}
  \newcommand{\CPLendCJK}{\end{CJK}}
\else
  \newcommand{\CPLbeginCJK}{}
  \newcommand{\CPLendCJK}{}
\fi

\begin{document}

\CPLbeginCJK\vspace* {-4mm} \begin{center}
\large\bf{\boldmath{NextCrystal: a Symmetry-Driven Generative Framework for \\ Crystal Structure Prediction}}
\footnotetext{\hspace*{-5.4mm}$^{*}$Corresponding authors. Emails: shiyin@iai.ustc.edu.cn; helx@ustc.edu.cn; zhu@iai.ustc.edu.cn.

\noindent\copyright\,{\cplyear}
\href{http://www.cps-net.org.cn}{Chinese Physical Society} and
\href{http://www.iop.org}{IOP Publishing Ltd}}
\\[5mm]
\normalsize \rm{}Jinming Mu$^{1}$, Lixin He$^{1,2*}$, Xudong Zhu$^{2*}$, and Shi Yin$^{2*}$
\\[3mm]\small\sl $^{1}$Laboratory of Quantum Information, University of Science and Technology of China, Hefei 230026, Anhui, China

$^{2}$Institute of Artificial Intelligence, Hefei Comprehensive National Science Center, Hefei 230088, Anhui, China
\\[4mm]\normalsize\rm{}(Received xxx; accepted manuscript online xxx)
\end{center}
\CPLendCJK
\vskip 1.5mm

\small{\narrower Crystal structure prediction (CSP), which aims to predict the 3D atomic arrangement of a crystal from its composition, is central to materials discovery and mechanistic understanding. Crystal symmetry plays a crucial role in CSP, but given the composition in a unit cell, existing methods either struggle with the NP-hard combinatorial challenge of enforcing symmetry rigorously or rely on retrieving known templates, inherently limiting both physical fidelity and the discovery of genuinely new materials. To address this challenge, we introduce NextCrystal, a symmetry-driven generative framework that employs large language models to encode chemical semantics and directly generate fine-grained Wyckoff site patterns from atomic stoichiometry, eliminating reliance on database lookups. To overcome the combinatorial complexity of site assignments, we incorporate domain knowledge via an efficient, linear-complexity heuristic beam search, rigorously enforcing algebraic consistency between site multiplicities and atomic stoichiometry. By integrating this symmetry-consistent template into a diffusion backbone, the framework constrains the stochastic generative trajectory to a physically plausible geometric manifold. NextCrystal achieves state-of-the-art performance on stability, uniqueness, and novelty (SUN) benchmarks, as well as superior structural matching, establishing a rigorous paradigm for exploring previously unexplored crystallographic space without relying on prior structural templates. As a representative application, first-principles screening of HfO$_2$ candidates generated by NextCrystal identifies a previously unreported dynamically stable \textit{Pnma} phase, 0.056~eV/atom lower in energy than the conventional high-pressure \textit{Pnma} phase.

\par}\vskip 3mm
\normalsize\noindent{\narrower{\textbf{Keywords:} AI for Science ; Materials Science ; Crystal Structure Prediction ; Space Group Symmetry ; Large Language Models ; Constrained Optimization}

\par}\vskip 2mm
\normalsize\noindent{\narrower{DOI: \href{http://dx.doi.org/10.1088/0256-307X/\cplvol/\cplno/\cplpagenumber}{10.1088/0256-307X/\cplvol/\cplno/\cplpagenumber}}

\par}\vskip 5mm
Crystalline materials are fundamental to modern technologies spanning energy, electronics, medicine, and aerospace, and advancing these sectors relies on identifying novel crystal structures with tailored properties \cite{green2014emergence,berger2020semiconductor,jiang2024rapid,ding2024amorphous,langer2004designing}. The vastness of the compositional and structural parameter space presents a fundamental bottleneck for materials discovery. Conventional pipelines, guided by chemical intuition and constrained by costly computations and experiments \cite{seh2017combining,marzari2021electronic}, are intrinsically inefficient at exploring this space at scale.

In recent years, artificial intelligence (AI), and deep learning-based generative models in particular, have emerged as a promising route to proposing candidate materials at scale, demonstrating superior generalization capabilities and significant advantages in computational efficiency \cite{jiao2023crystal,antunes2024crystal,jiaospace,NEURIPS2024_447d012b,ZHU20243469,han2025efficient,zeni2025generative}. Within this landscape, the crystal structure prediction (CSP) task \cite{jiao2023crystal,jiaospace,han2025efficient}, which seeks the three-dimensional arrangement of atoms in the unit cell from a given composition, plays a pivotal role in AI-driven materials design workflows \cite{NEURIPS2024_447d012b}. Diffusion models have become the dominant architecture for 3D structure generation \cite{xu2023dream3d,wang2025diffusion}, and, for the CSP task, diffusion-based approaches also deliver state-of-the-art performance \cite{jiaospace,zeni2025generative}.

However, the CSP task is fundamentally governed by stringent crystallographic rules, imposing rigid constraints on the design methodology. Physically plausible crystal structures typically obey the symmetry constraints imposed by specific space groups, which strictly constrain the allowed Wyckoff positions, determining atomic multiplicities, site symmetries, and the relative arrangement of atoms within the unit cell. Only when the given chemical composition is mathematically consistent with the corresponding Wyckoff positions can physically reasonable crystal structures be obtained.

\begin{figure*}[t]
	\centering
	\includegraphics[scale=0.64]{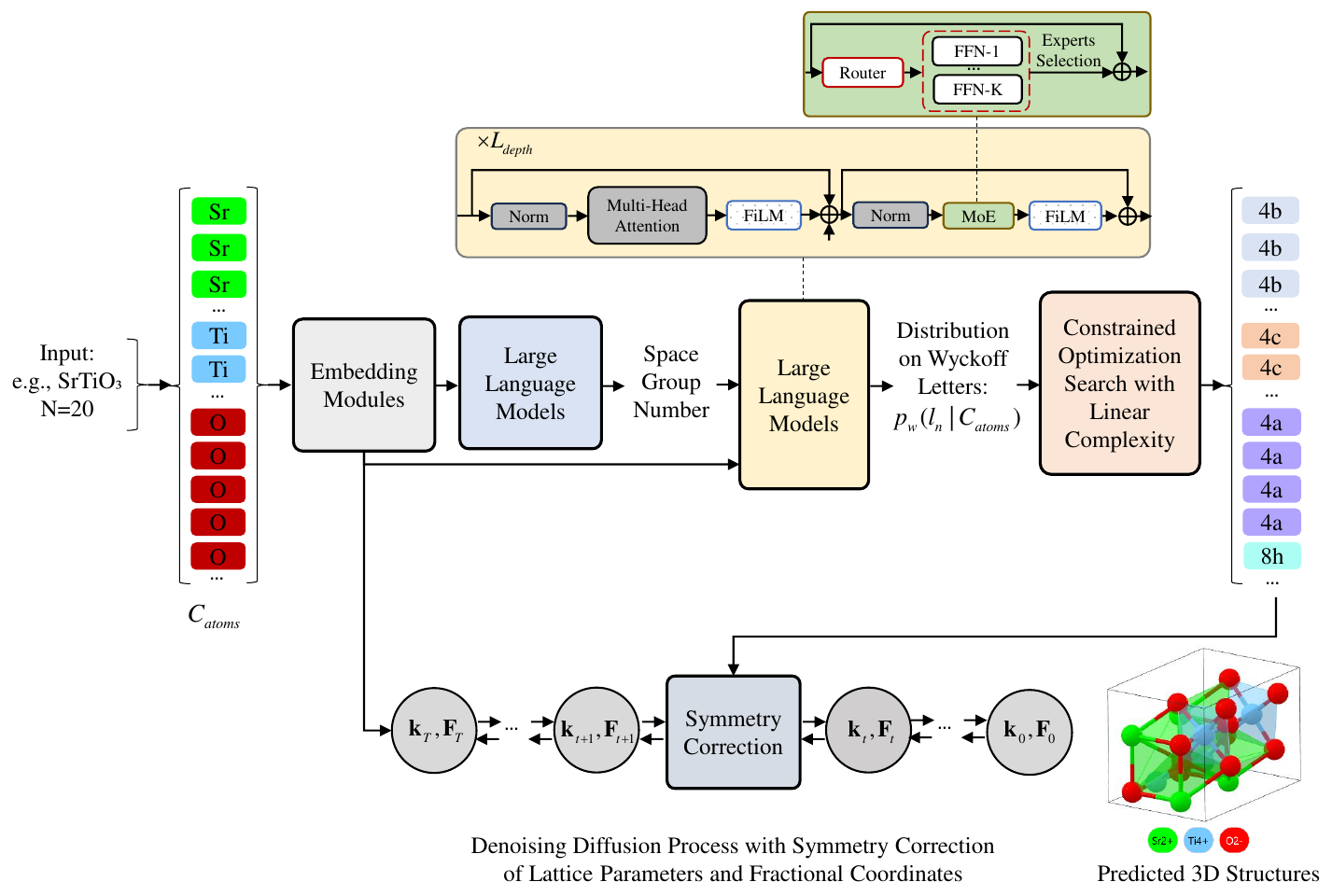}
	\caption{Schematic of the proposed symmetry-driven generative framework.}
	\label{framework}
\end{figure*}

However, rigorously enforcing these physical constraints imposed by space groups remains a universal bottleneck across the entire CSP domain.
Determining an exact combination of Wyckoff positions whose multiplicities precisely match the atom counts of a given unit-cell composition is an NP-hard combinatorial optimization problem with an exponentially growing search space.
Existing methods struggle to efficiently resolve this task, often requiring partial compromises. For instance, Xin et al.~\cite{xin2025symmetry} proposes a pruning-based search for Wyckoff positions to address site assignments. Nevertheless, the approach remains fundamentally a brute-force enumeration. As the number of atoms in a unit cell increases, the possible partitions of atoms into available Wyckoff sites grow exponentially.
For large systems, exhaustive enumeration becomes computationally prohibitive. Moreover, lacking a likelihood-based ranking mechanism, such methods cannot efficiently identify the most physically plausible candidates from the massive, unguided pool of combinations.

Within the landscape of generative AI, existing methods only partially address the discrete combinatorial challenge of site assignments. Early 3D generative models, such as CDVAE \cite{DBLP:conf/iclr/XieFGBJ22} and DiffCSP \cite{DBLP:conf/nips/Jiao0LHCLL23}, bypass this challenge by operating directly on continuous coordinates. These models rely on neural networks to implicitly learn crystallographic symmetries, leaving the generative trajectory unconstrained in a high-dimensional search space.
CrystaLLM \cite{antunes2024crystal} adopts a text-generation approach, autoregressively predicting Crystallographic Information File (CIF)  formats; however, any generated symmetry or Wyckoff symbols are purely probabilistic textual outputs, lacking explicit algebraic or physical constraints.
 Some recent methods condition the generator on a space group label \cite{NEURIPS2024_447d012b,zeni2025generative}, but they do not explicitly represent or enforce Wyckoff positions, capturing symmetry only in a coarse, global sense.

Moving beyond coarse global symmetry, several recent AI methods attempt to explicitly utilize and strictly enforce Wyckoff positions, yet they remain hindered by the intractable complexity of exact site assignments. While WyCryst \cite{ZHU20243469} elegantly guides generation via Wyckoff-based latent representations, its soft-penalty sampling cannot guarantee strict stoichiometric alignment. Consequently, extracting 3D structures consistent with Wyckoff positions relies on post-hoc filtering, a process that can become computationally prohibitive given the NP-hard nature of combinatorial site assignments. CrystalFormer \cite{DBLP:journals/corr/abs-2403-15734} presents a compelling framework representing crystals as Wyckoff sequences with logit masking for symmetry enforcement. While effective for generating structures within a given space group number, the final stoichiometry is an emergent outcome of the sampling process rather than a hard constraint. Since the model focuses on generative modeling, it does not employ an explicit constrained optimization mechanism to align site multiplicities with a given composition and atomic number in a cell. DiffCSP++ \cite{jiaospace} goes further by enforcing strict space group symmetry defined by Wyckoff templates during the 3D material generation process. It employs two strategies for obtaining these templates: the first directly utilizes the ground-truth Wyckoff templates of the target material to guide structure generation. However, in practical CSP scenarios, these ground-truth templates are inherently unknown a priori. As an alternative, it relies on a retrieval-based approach, which assumes that a suitable space group and Wyckoff template are already known in the given lookup database. In practice, these templates are retrieved from existing structures via metric learning methods, such as CSPML \cite{KUSABA2022111496}. However, restricting generation to retrieved templates confines the model to known symmetry patterns, thereby limiting its ability to discover genuinely new space-group and Wyckoff configurations for a given composition.

To address these fundamental limitations, we propose an efficient symmetry-driven generative framework, named NextCrystal, which enables \textit{ab initio} generation of fine-grained Wyckoff site assignments directly from composition and atom counts, independent of any structural priors, as illustrated in Fig.~\ref{framework}. Building upon the probabilistic distributions inferred by large language models (LLMs), we explicitly overcome the intractable exponential complexity of combinatorial site assignments discussed above by designing an efficient heuristic constrained-optimization search. This algorithm drastically compresses the time complexity from an exponential scale down to a linear scale, rendering the search computationally tractable while rigorously enforcing algebraic consistency between site multiplicities and atomic stoichiometry without relying on computationally prohibitive brute-force enumeration. Crucially, the predicted symmetries are then imposed as hard geometric constraints to guide and modulate the diffusion process for three-dimensional crystal geometries, thereby rectifying the denoising trajectory, confining the generative process to a physically valid manifold, guaranteeing both structural plausibility and computational precision.

Below, we outline the core methodology of NextCrystal in this Section. For full implementation details, as well as training and testing configurations, please refer to Appendix A and B, respectively.

First, we use two large language models (LLMs) to directly infer fine-grained crystallographic symmetry from compositional inputs.
Both models are built upon the Transformer architecture \cite{DBLP:conf/nips/VaswaniSPUJGKP17}, and we replace the standard feed-forward networks within each Transformer block with soft mixture-of-experts (SoftMoE) layers \cite{puigcerversparse} to enhance model capacity. This design enables multiple experts to specialize in distinct compositional patterns, substantially boosting model capacity without a commensurate rise in computational cost, as only a small, softly weighted subset of experts is activated at each position.

The first LLM, denoted as $LLM_g$, takes as input an atomic sequence representation of the composition, expanded to explicitly reflect the atom counts $N$ in the unit cell, and outputs a probability distribution over the 230 crystallographic space groups:
\begin{equation}
\label{llm1}
S_{g} =  {\arg\max}_i \, p_{g}(i|C_{\text{atoms}}), \,\,\, p_{g}(i|C_{\text{atoms}}) = LLM_g(C_{\text{atoms}}), \,\,\, 1 \leq i \leq 230,
\end{equation}
where \( C_{\text{atoms}} = [A,A,A,\dots,B,B,B,B,B,\dots,C,C,\dots] \) is the expanded atomic sequence for elements $ A, B, C, \dots $ derived from the given composition, proportionally expanded to match the number of atoms in the unit cell; \( p_{g}[i] \) is the predicted probability that the crystal belongs to the $i$-th space group; $S_{g}$ represents the predicted space group.

The second LLM, denoted as $LLM_w$, takes the same expanded atomic sequence together with the predicted space group $S_g$ and outputs a probability distribution over Wyckoff letters compatible with that space group for each atom. Rather than concatenating $S_g$ naively to the input, we encode the predicted space group into a continuous embedding and inject it into each transformer block via the feature-wise linear modulation (FiLM)  \cite{perez2018film} module. This FiLM-based conditioning ensures that Wyckoff letter predictions are explicitly modulated by the inferred space group, rather than being learned in an implicitly entangled fashion. Concretely, FiLM($S_g$) produces per-channel scaling and shifting coefficients that modulate the intermediate representations of $LLM_w$, allowing the model to adapt its representation to the symmetry class:
\begin{equation}
	\label{llm2}
p_{w}(l_n=l|C_{\text{atoms}}) = LLM_w\big(C_{\text{atoms}},  \mathrm{FiLM}(S_{g})\big),  \quad l_n \in L(S_{g}),\quad 1 \leq n \leq N,
\end{equation}
where $p_{w}(l_n=l|C_{\text{atoms}})$ denotes the probability that the $n$-th atom is assigned to letter $l$, $L(S_{g})$ is the set of admissible Wyckoff letters for space group $S_{g}$.  Specific details regarding the network architectures and hyperparameter configurations of these two large language models are provided in Appendix A.1-A.2 and B.

Second, we formulate the Wyckoff letter assignment for each atom as a constrained optimization problem based on the probability distribution $ p_{w}(l_n|C_{\text{atoms}}) $ predicted by the LLM, and propose an efficient algorithm to solve it. This formulation strictly adheres to the combinatorial constraints imposed by the Wyckoff site multiplicity rules under the inferred space group symmetry. Specifically, the constrained maximization objective is defined as follows:
\begin{equation}
	\label{constraint}
	\begin{aligned}
		\{l_n^*\}_{n=1}^N = & \, \arg\max_{\{l_n\}_{n=1}^N} \quad \sum_{n=1}^{N} \log p_{w}(l_n|C_{\text{atoms}}), \\
		s.t. \,\,\, &  l_n \in L(S_g),  \\
		& \text{Count}_e(l_n) \bmod \text{mult}(l_n)  =0,\\
		& \sum_{l \in L(S_g)} \text{Count}_e(l)  = N_e, \\
		& \forall\, e \in \mathcal{E}=\{A, B, C, \dots\},
	\end{aligned}
\end{equation}
where \(\text{mult}(l_n)\) refers to the multiplicity of Wyckoff letter \(l_n\),  and \(\text{Count}_e(l_n)\) is the number of atoms of element \(e\) assigned to \(l_n\), \(N_e\) is the number of atoms of element \(e\) in a cell, the set \(\{l_n^*\}_{n=1}^N\) represents the optimal Wyckoff letter assignment. The key point here is that $\text{Count}_e(l_n)$ must be an integer multiple of $\text{mult}(l_n)$, a hard constraint that cannot be satisfied by simply using the maximum probability outputs of each token from the LLM. Fundamentally, deriving the exact global optimum under these precise constraints constitutes an NP-hard combinatorial optimization problem. Consequently, finding the solution for Eq. (\ref{constraint}) via brute-force enumeration is computationally intractable, as the search space grows exponentially with the number of atoms, scaling as $O(|L(S_g)|^N)$. To overcome this bottleneck, we propose a beam search algorithm to efficiently solve the optimization problem. By maintaining only the most promising candidates within a fixed beam width, this strategy drastically compresses the time complexity from exponential to a linear $O(N)$, rendering the prediction of physically valid Wyckoff assignments computationally tractable.  Comprehensive details regarding this algorithm are provided in Appendix A.3.

To generate the 20 candidates for each test composition, we employ a hierarchical sampling strategy to ensure both macroscopic and microscopic structural diversity. First, we select the Top-5 most probable space groups predicted by $LLM_g$. Subsequently, for each selected space group, we extract the Top-4 valid Wyckoff assignments using our constrained beam search algorithm. This combination yields exactly 20 distinct symmetry templates per composition, which are then used to guide the diffusion generation process.

Third, we guide and constrain the diffusion-based generation of 3D atomic structures using the predicted space group and Wyckoff site template.  To instantiate this approach, we adopt the graph neural network (GNN) applied by DiffCSP++\cite{jiaospace} as the backbone network of diffusion model, effectively repurposing its exploration to strictly adhere to the symmetry constraints inferred by our method. Specifically, we employ a joint-rectification mechanism on both lattice and fractional coordinates. For the lattice, we constrain its $O(3)$-invariant logarithmic parameterization by applying a binary mask dictated by the inferred crystal family, ensuring the unit cell geometry strictly adheres to the specific constraints on lattice lengths and inter-axial angles defined by the crystal family. Simultaneously, for the fractional coordinates $\mathbf{F}_{t}$, we employ a subspace projection and reconstruction mechanism, effectively rectifying any symmetry violations in each denoising state of fractional coordinates, ensuring that the generated crystal structure strictly resides on the geometric manifold defined by the target Wyckoff templates. Further details regarding the lattice masking and coordinate reconstruction are available in Appendix A.4.

To evaluate our approach, we benchmark it against a comprehensive suite of generative models. We adopt DiffCSP++ \cite{jiaospace} as the primary baseline, as it represents the current state-of-the-art (SOTA) for standard CSP and serves as the most stringent benchmark for composition-conditioned structural generation. To provide a broader context for structural reconstruction fidelity, we also encompass CDVAE \cite{DBLP:conf/iclr/XieFGBJ22}, DiffCSP \cite{DBLP:conf/nips/Jiao0LHCLL23}, and CrystaLLM \cite{antunes2024crystal} in our comparative analysis, specifically regarding the Matching Rate metric. While MatterGen \cite{zeni2025generative} and CrystalFormer \cite{DBLP:journals/corr/abs-2403-15734} are leading methods in materials generation, they are excluded from direct quantitative comparison in this work. These models primarily operate on a joint generation paradigm that co-samples chemical compositions and structural coordinates. Adapting them into dedicated CSP models and training them from scratch without the original authors' optimization protocol risks an unfair and unrepresentative comparison. Thus, a direct evaluation is omitted.

To provide a comprehensive evaluation of our model, we adopt the Top-20 SUN (stability, uniqueness, and novelty) metrics based on the core principles of \cite{zeni2025generative}, alongside the Top-20 Matching Rate \cite{DBLP:conf/nips/Jiao0LHCLL23}. The detailed definitions and adapted calculation protocols for these metrics, designed to accommodate our extensive Top-20 evaluation setting, are provided in Appendix C.1 and C.2. The SUN metrics emphasize physical plausibility, structural diversity, and originality, aspects that are more directly aligned with the objectives of open-ended discovery in materials science; while the Matching Rate prioritizes structural fidelity by quantifying the precision with which the model reconstructs the experimental ground-truth (GT) geometry within defined coordinate and symmetry tolerances. It is worth noting that these two metrics can be inherently contradictory in a Top-1 evaluation because a model must choose between replicating a known structure to achieve a high Matching Rate or exploring a previously unknown one to achieve high Novelty. This tension is effectively resolved in our Top-20 setting. By generating multiple candidates for each composition, the model is afforded the capacity to simultaneously demonstrate structural fidelity by recovering the experimental ground truth and chemical creativity by proposing stable, novel phases. Thus, reporting both metrics at Top-20 provides a holistic assessment of the model's dual role as both a precise structural reconstructor and a robust discovery engine for uncharted materials space.

Our experiments are conducted on three standard CSP benchmark datasets: MP-20 \cite{jain2013commentary}, Perov-5 \cite{castelli2012new},  and MPTS-52 \cite{jiaospace}. These datasets were previously used in DiffCSP++ and are employed here with identical training, validation and testing splits to ensure a fair comparison. Detailed descriptions and statistical information regarding these three datasets are available in Appendix C.2. The experimental results for the DiffCSP++ baseline are obtained using its official open-source scripts for standard training and inference \cite{diffcsppp_note}. Table \ref{tab:sun_results_vertical} presents a side-by-side comparison between the performance of this original method and that of our symmetry-enhanced method.

\begin{table*}[tb]
	\centering
	\caption{Comparison of Top-20 SUN metrics (\%) across three benchmarks. Results for DiffCSP++ \cite{jiaospace} are reproduced using the official source code implementation \cite{diffcsppp_note}, utilizing the CSPML \cite{KUSABA2022111496} retrieval method to obtain space group templates as described in the original study.}
	\label{tab:sun_results_vertical}
	\vspace{0.5em}
	\begin{tabular}{ll|cccc}
		\toprule
		Dataset & Method & Stability & Uniqueness & Novelty & \textbf{SUN} \\
		\midrule
		\multirow{2}{*}{MP-20}
		& DiffCSP++ & 36.51 & \textbf{100.0} & 54.82 & 14.55 \\
		& Ours & \textbf{81.79} & \textbf{100.0} & \textbf{93.67} & \textbf{35.58} \\
		\midrule
		\multirow{2}{*}{Perov-5}
		& DiffCSP++ & 75.77 & \textbf{100.0} & 99.52 & 73.39 \\
		& Ours & \textbf{94.98} & \textbf{100.0} & \textbf{99.67} & \textbf{92.42} \\
		\midrule
		\multirow{2}{*}{MPTS-52}
		& DiffCSP++ & 11.95 & \textbf{100.0} & 45.97 & 5.44 \\
		& Ours & \textbf{42.43} & \textbf{100.0} & \textbf{70.53} & \textbf{25.94} \\
		\bottomrule
	\end{tabular}
\end{table*}

\begin{table*}[tb]
	\centering
	\caption{Comparison of Top-20 Matching Rates (\%) across three benchmarks. Experimental results for CDVAE \cite{DBLP:conf/iclr/XieFGBJ22}, DiffCSP \cite{DBLP:conf/nips/Jiao0LHCLL23}, and CrystaLLM \cite{antunes2024crystal} are taken from Antunes et al.~\cite{antunes2024crystal}. Results for DiffCSP++ \cite{jiaospace} are reproduced from the official source code \cite{diffcsppp_note} using the CSPML \cite{KUSABA2022111496} retrieval-based template method as described in the original study.}
	\label{tab:match_rate_results}
	\begin{tabular}{ll|c}
		\toprule
		Dataset & Method & Matching Rate \\
		\midrule
		\multirow{5}{*}{MP-20}
		& CDVAE & 66.95 \\
		& DiffCSP & 77.93 \\
		& CrystaLLM & 73.97 \\
		& DiffCSP++ & 74.65 \\
		& Ours & \textbf{81.70} \\
		\midrule
		\multirow{5}{*}{Perov-5}
		& CDVAE & 88.51 \\
		& DiffCSP & 98.60 \\
		& CrystaLLM & 97.60 \\
		& DiffCSP++ & 99.84 \\
		& Ours & \textbf{99.95} \\
		\midrule
		\multirow{5}{*}{MPTS-52}
		& CDVAE & 20.79 \\
		& DiffCSP & 34.02 \\
		& CrystaLLM & 33.75 \\
		& DiffCSP++ & 40.53 \\
		& Ours & \textbf{43.26} \\
		\bottomrule
	\end{tabular}
\end{table*}

As shown in Table~\ref{tab:sun_results_vertical}, our method yields notable and consistent improvements over DiffCSP++ across all three benchmarks on the comprehensive SUN metrics.
On the  MP-20 dataset, our method improves stability and novelty by approximately 124\% and 71\% respectively relative to the baseline. The advantages are even more pronounced on the challenging MPTS-52 dataset, where our method achieves a relative gain of about 255\% in stability and 53\% in novelty. Crucially, these component-level improvements compound to yield a striking 376\% increase in the overall SUN metric. Even on Perov-5, which features relatively simple structural motifs, our method maintains a clear advantage in stability while matching the saturated uniqueness and novelty scores of the baseline. The significant rise in the overall SUN metric across all benchmark datasets confirms that our method generates candidates that are simultaneously more physically stable and structurally novel, while preserving the high standards of uniqueness.

\begin{figure*}
	\centering
	\includegraphics[width=\textwidth]{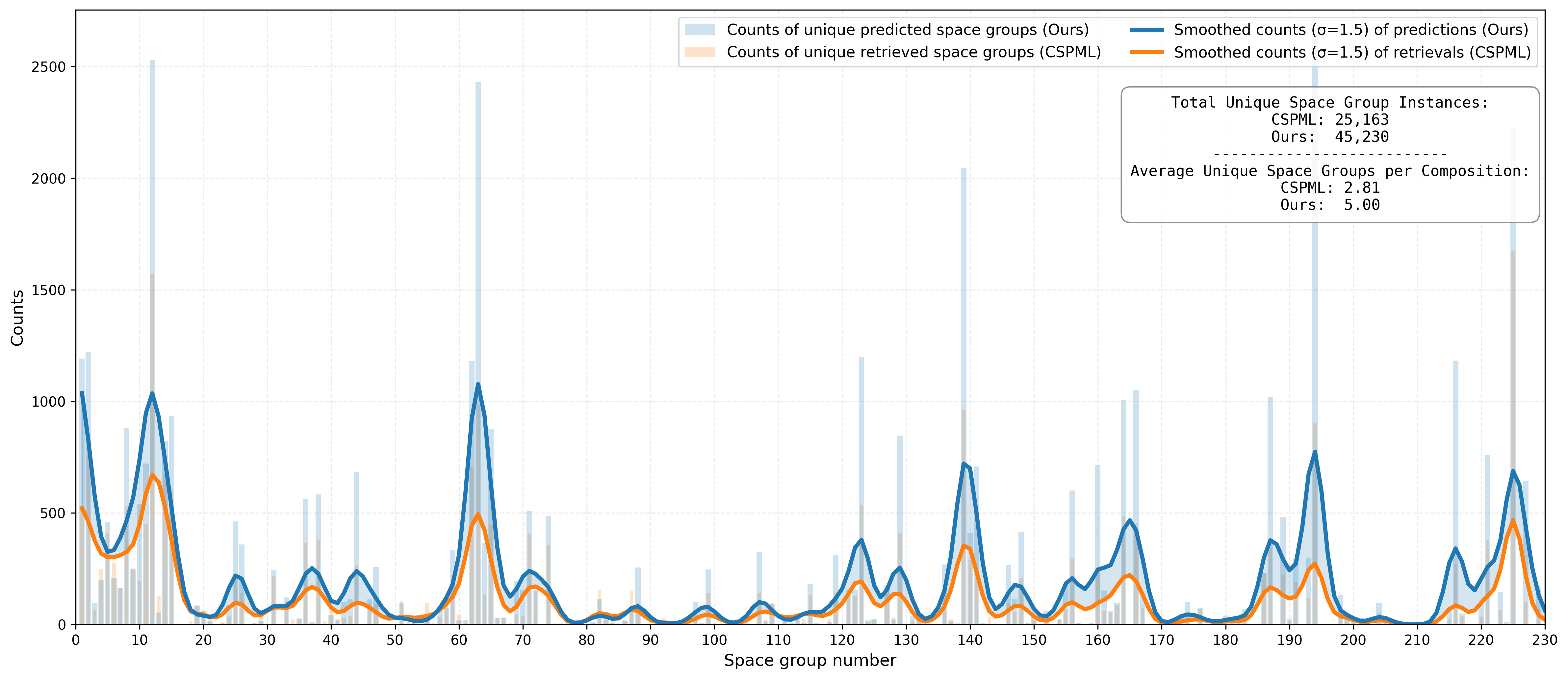}
	\caption{Absolute frequency distribution (raw counts) of space groups among the candidates across the testing set of MP-20. Counts are calculated using a unique-counting principle per composition, where any specific space group present in the candidate structures for a single testing composition is counted exactly once. Our generative approach yields a profoundly larger total unique count, demonstrating a consistently higher richness in symmetry exploration per composition compared to the localized redundancy of the retrieval baseline, i.e. CSPML \cite{KUSABA2022111496}.}
	\label{fig:sg_distribution}
\end{figure*}

To gain deeper insights into the distinct generative behaviors of retrieval-based and symmetry-driven methods, we conducted detailed case studies on two representative samples, Zr$_3$Fe$_8$Mo and NaCr$_4$O$_8$, as provided in Appendix D (Fig. D.1 - Fig. D.2). An analysis of these examples reveals that the baseline systematically struggles to satisfy the critical criteria for successful discovery. It often either relies on an exact structural match of a documented phase, which guarantees stability but completely fails the novelty criterion, or forces the atoms into a retrieved but topologically mismatched template. In the latter case, atoms are coerced into energetically unfavorable coordination environments to satisfy incorrect symmetry constraints, directly leading to poor stability and severe microscopic strain. In contrast, our symmetry-driven approach discovers novel, stable phases by adaptively allocating atoms to higher-multiplicity orbits with critical geometric flexibility. This transition from retrieval-based memorization to generative reasoning allows NextCrystal to yield candidates that successfully satisfy all three SUN criteria.

The performance disparity observed in Table~\ref{tab:sun_results_vertical} and Appendix D (Fig. D.1 - Fig. D.2) stems from the fundamental transition from retrieval-based memorization to generative reasoning. The baseline relies on CSPML, which essentially operates as a static lookup system. It rests on the isomorphism assumption, presupposing that for any new material, a structurally identical twin already exists in the training database. Consequently, this approach fails in data-sparse regimes or when exploring truly novel materials where no such prototype exists. When the model forces a new querying composition into a retrieved but topologically mismatched template, it introduces geometric frustration. The atoms are coerced into energetically unfavorable coordination environments to satisfy incorrect symmetry constraints, directly leading to poor stability. This reliance on data lookup inherently confines the model to the known structural motifs, rendering it incapable of discovering new polymorphs that lie outside the existing library.

In stark contrast, our method reformulates symmetry determination as \textit{ab initio} generative inference coupled with rigorous constrained optimization. Instead of retrieving existing solutions, the model learns a unified representation space from diverse structural data, enabling it to generalize and derive valid Wyckoff patterns for unseen compositions based on chemical rules. Crucially, a heuristic combinatorial optimization algorithm enforces algebraic consistency, ensuring the sum of site multiplicities matches the atomic stoichiometry. This mechanism empowers the model to explore uncharted chemical space while confining stochastic diffusion to a symmetry-compliant configurational space, guaranteeing output rationality. By eliminating symmetry-breaking artifacts of unconstrained generation, our framework enables inverse design of materials with high structural novelty and strict physical rigor.

To further quantify generative symmetry diversity, we evaluate the distribution of unique space group instances generated across the testing set, as shown in Fig. \ref{fig:sg_distribution}. To faithfully capture the true exploratory richness for each individual composition, we apply a strict unique-counting principle: a unique space group instance is defined such that if a specific space group appears within the candidate structures generated for a single testing composition, it is counted exactly once for that composition, regardless of its redundancy. The results demonstrate that the CSPML retrieval baseline \cite{KUSABA2022111496} often collapses into retrieving multiple structural templates that share the identical space group for a given composition. Because of this localized redundancy, multiple retrieved candidates frequently contribute only a single unique space group count, which drastically shrinks its total aggregate score. In contrast, our generative approach explicitly infers multiple mutually distinct space groups directly from the probability distribution predicted by the language models. This mechanism yields a profoundly larger total count of unique instances and demonstrates a consistently richer and more abundant exploration across  different space groups for each of the composition, as evidenced by the higher frequency observed across nearly all space group indices in Fig. \ref{fig:sg_distribution}.

Complementing the discovery-oriented SUN metrics, Table \ref{tab:match_rate_results} presents the Top-20 Matching Rate, which quantifies the model's ability to reconstruct experimentally known ground-truth structures. As shown, our method consistently achieves state-of-the-art performance across all three benchmarks, outperforming all of the baseline methods. This result is scientifically significant when interpreted in the context of generative design. In \textit{ab initio} materials discovery, the Matching Rate confirms that the model's generative distribution covers the manifold of physically realized structures.
A high Matching Rate indicates that the symmetry constraints imposed by our framework do not overly restrict the search space or exclude valid ground-truth configurations. Instead, they effectively guide the diffusion process within a chemically reasonable manifold. Consequently, the combination of superior SUN scores (Table \ref{tab:sun_results_vertical}) and leading Matching Rates (Table \ref{tab:match_rate_results}) demonstrates that our method resolves the trade-off between exploration and exploitation: it is robust enough to recover known stable phases with high fidelity while possessing the generative flexibility to propose novel, stable polymorphs beyond the existing samples.

To demonstrate the practical utility of our model, we applied the MP-20 pre-trained NextCrystal model to the generation of HfO$_2$ structures. HfO$_2$ exhibits a rich polymorphic landscape, with closely competing nonpolar, ferroelectric, and antiferroelectric-related phases, making the identification of energetically accessible crystal structures central to understanding and tuning its functional properties \cite{Zhu2024}. We use (HfO$_2$)$_n$ ($n = 1$--$12$) as the initial inputs to NextCrystal. For each input, NextCrystal generates up to 100 structures by sampling 20 space groups and 5 sets of Wyckoff position assignments, resulting in a total of 887 HfO$_2$ candidate structures. These generated structures are subsequently validated using first-principles calculations within a workflow integrating ABACUS \cite{Li2016,Chen2010}, PYATB \cite{Jin2023}, ZStar, and PHONOPY \cite{phonopy-phono3py-JPCM} packages. The exchange--correlation energy is treated within the PBEsol functional \cite{Perdew2008}, and numerical atomic-orbital basis sets of TZDP and DZP were employed for Hf and O, respectively \cite{Lin2021}.

\begin{figure*}
	\centering
	\includegraphics[width=\textwidth]{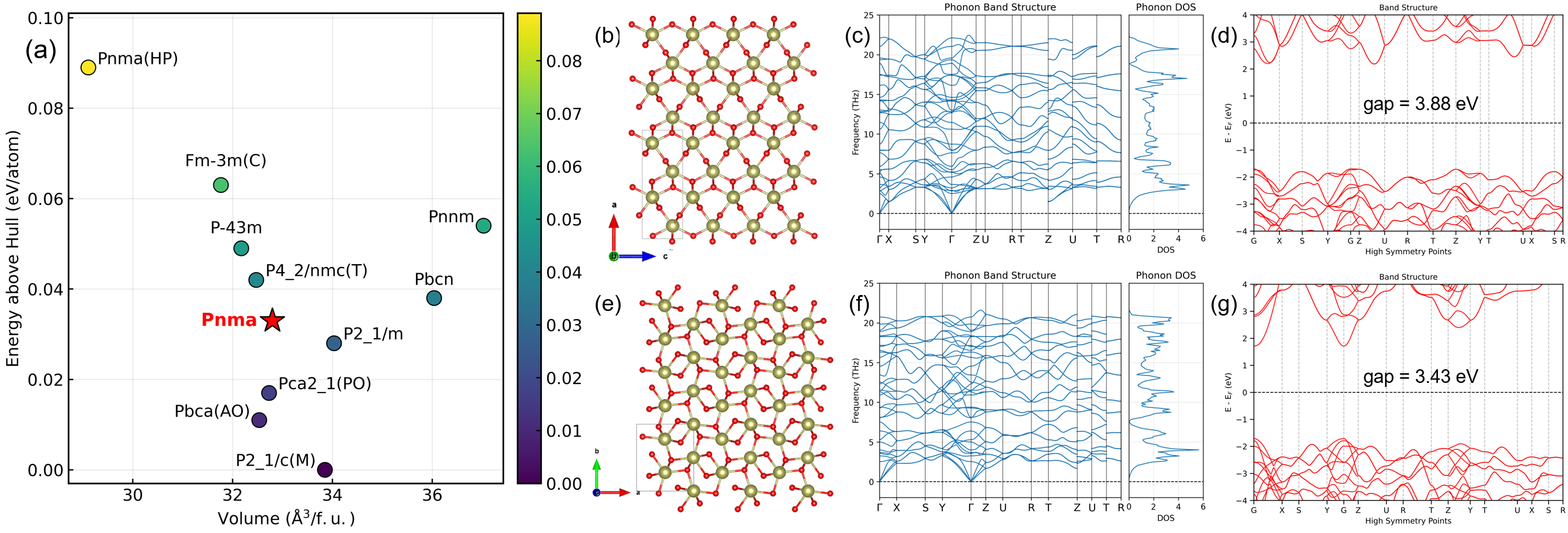}
\caption{Comparison between the newly identified \textit{Pnma} phase of HfO$_2$ and the conventional high-pressure (HP) \textit{Pnma} phase. (a) Distribution of the 11 stable HfO$_2$ phases in the volume--energy space, where the star marks the newly identified \textit{Pnma} phase. (b)--(d) Crystal structure, phonon dispersion, and electronic band structure of the newly identified \textit{Pnma} phase. (e)--(g) Crystal structure, phonon dispersion, and electronic band structure of the HP \textit{Pnma} phase.}
	\label{fig:hfo2_pnma}
\end{figure*}

\begin{table*}[htbp]
	\centering
	\footnotesize
	\caption{Symmetry-inequivalent atomic sites, fractional coordinates, and Born effective charge tensors $Z^{*}$ of the newly identified \textit{Pnma} phase of HfO$_2$.}
	\label{tab:hfo2_pnma_bec}
	\setlength{\tabcolsep}{4pt}
	\renewcommand{\arraystretch}{1.25}
	\begin{tabular}{lll}
		\toprule
		Wyckoff site & Fractional coordinates & Born effective charge tensor $Z^{*}$ ($e$) \\
		\midrule
		4c (Hf) & $(0.3817,\ 0.7500,\ 0.2524)$ &
		\BECTensor
		{5.008}{0.000}{0.314}
		{0.000}{5.238}{0.000}
		{0.130}{0.000}{4.999}
		\\[8pt]
		
		4c (O1) & $(0.7152,\ 0.7500,\ 0.5778)$ &
		\BECTensor
		{-2.664}{0.000}{0.870}
		{0.000}{-3.016}{0.000}
		{0.812}{0.000}{-2.046}
		\\[8pt]
		
		4c (O2) & $(0.9746,\ 0.7500,\ 0.7391)$ &
		\BECTensor
		{-2.346}{0.000}{-0.068}
		{0.000}{-2.223}{0.000}
		{0.000}{0.000}{-2.954}
		\\
\bottomrule
\end{tabular}
\end{table*}
	
After structural relaxation, duplicate removal, and stability assessment, 11 candidates remain as both stable and structurally distinct. Their distribution in the volume--energy space is shown in Fig.~\ref{fig:hfo2_pnma}(a). Among them, we identify a new \textit{Pnma} polymorph (space group No.~62) that, to our knowledge, has not been reported previously. Relative to the representative HfO$_2$ phases, namely monoclinic P2$_1$/c (M), antiferroelectric orthorhombic Pbca (AO), polar orthorhombic Pca2$_1$ (PO), and tetragonal P4$_2$/nmc (T), the new \textit{Pnma} phase occupies an intermediate position within the low-energy metastable manifold. Its energy above hull is 0.033~eV/atom, which is higher than those of M, AO, and PO by 0.033, 0.022, and 0.016~eV/atom, respectively, but lower than that of T by 0.009~eV/atom. Its intermediate energy and volume suggest that it may serve as a metastable phase connecting the AO, PO, and T phases.

The newly identified \textit{Pnma} polymorph [Fig.~\ref{fig:hfo2_pnma}(b)] is structurally distinct from the conventional high-pressure (HP) \textit{Pnma} phase of HfO$_2$ [Fig.~\ref{fig:hfo2_pnma}(e)]. It is 0.056~eV/atom lower in energy than the HP phase, and its phonon spectrum exhibits no imaginary frequencies, confirming its dynamical stability [Fig.~\ref{fig:hfo2_pnma}(c)].

This phase also exhibits distinctive electronic and dielectric properties. It has a band gap of 3.88~eV [Fig.~\ref{fig:hfo2_pnma}(d)], 0.45~eV larger than that of the HP phase (3.43~eV) [Fig.~\ref{fig:hfo2_pnma}(g)]. Although this \textit{Pnma} phase is centrosymmetric and therefore nonpolar, its Born effective charge tensors still exhibit pronounced anomalous dynamical charge response, indicating substantial Hf--O hybridization. The symmetry-inequivalent atomic positions and corresponding Born effective charge tensors are summarized in Table~\ref{tab:hfo2_pnma_bec}. The calculated static dielectric constants, $\varepsilon_{xx}=17.85$, $\varepsilon_{yy}=19.93$, and $\varepsilon_{zz}=18.05$, reveal moderate dielectric anisotropy and substantial lattice polarizability.

This representative case shows that the present model can identify physically meaningful metastable or stable oxide phases in previously overlooked low-energy regions of configurational space. Notably, the model is not trained specifically for the HfO$_2$ system, underscoring its strong generalization capability.

In conclusion, this work introduces NextCrystal, a symmetry-driven CSP framework that redefines crystal structure prediction by shifting the paradigm from retrieval-based memorization to constrained \textit{ab initio} generative inference. 
By marrying large language models with a linear-complexity heuristic search, NextCrystal successfully renders the inherently NP-hard combinatorial challenge of symmetry enforcement computationally tractable. 
This approach yields exceptional performance in both data-sparse regimes and the exploration of novel material spaces, outperforming state-of-the-art baselines across comprehensive SUN metrics and matching rates. 
Crucially, the discovery and first-principles validation of a new, dynamically stable \textit{Pnma} phase in the HfO$_2$ system highlights NextCrystal's unique capability to uncover promising candidates entirely beyond known structural prototypes.

Building upon this foundation, future extensions will integrate high-fidelity AI electronic structure prediction models, such as TraceGrad \cite{DBLP:conf/icml/YinPWH25} and NextHAM \cite{yin2026advancing} developed by our group, to enable rapid, high-throughput screening of the generated candidates. 
Furthermore, subsequent iterations of this framework will pursue direct integration with high-throughput experimental workflows and automated synthesis laboratories. 
Establishing an active, closed-loop experimental feedback system where computational predictions are iteratively synthesized, characterized, and utilized to fine-tune the generative model is essential. 
This seamless alignment of computational design, AI-driven screening, and empirical validation will ultimately realize a fully autonomous and self-optimizing materials discovery pipeline.

\textit{Conflict of Interest Statement.} The authors declare that they have no conflict of interest.

\textit{Acknowledgements.} This work was supported by the Advanced Materials--National Science and Technology Major Project (Grant No. 2025ZD0618401), the National Natural Science Foundation of China (Grant No. 62506112, 12134012), the Strategic Priority Research Program of the Chinese Academy of Sciences (Grant No. XDB0500201), and the Anhui Provincial Science and Technology Breakthrough Project (Grant No. 202523o09050015). The numerical calculations were performed on the USTC High-Performance Computing facilities and Hefei advanced computing center.

\end{document}